\documentclass[twoside]{ilcws07}
\usepackage[latin1]{inputenc}
\usepackage[dvips]{graphicx,epsfig,color}
\usepackage{wrapfig,rotating}
\usepackage{amssymb,amsmath,array}

\def\KL{K\"all\'en-Lehmann}
\def\Journal#1#2#3#4{{#1} {\bf #2}, #3 (#4)}

\def\PLB{{\em Phys. Lett.}  B}
\def\PRL{\em Phys. Rev. Lett.}
\def\PRD{{\em Phys. Rev.} D}

\def\be{\begin{equation}}
\def\ee{\end{equation}}
\def\bea{\begin{eqnarray}}
\def\eea{\end{eqnarray}}

\begin{document}
\title{
HEIDI and the unparticle} 
\author{J.~J.~van~der~Bij and S.~Dilcher
\vspace{.3cm}\\
Albert-Ludwigs Universit\"at Freiburg - Institut f\"ur Physik \\
H. Herderstr. 3, 79104 Freiburg i.B. - Deutschland
}

\maketitle

\begin{abstract}
We compare the HEIDI models with the unparticle models.
We show that the unparticle models are a limiting case of the 
HEIDI models. We discuss consistency conditions.
\end{abstract}

\section{HEIDI models}
Following a basic construction introduced with the late Alfred Hill~\cite{hill}
the so-called HEIDI (High-D for higher dimension, in exchange for SUSY) models
were recently introduced~\cite{bij2006,dilcher,lorca} and discussed in a number of
conferences~\cite{thooft,moriond,lcws}. This is a new class of renormalizable models
that consist of an extension of the standard model with singlet fields.
The singlet fields can live in higher dimensions $d$, even fractional ones.
Nonetheless the theory stays renormalizable
when $d<6$, because one can have terms
in the Lagrangian that would be super-renormalizable in four dimensions.
Since the singlet sector cannot be probed directly, the experimental signature
shows up through the mixing of the standard model fields and the singlet fields.
As a consequence the standard model fields $\sigma$, that normally contain a 
single particle peak,
will satisfy a more general \KL\,\, spectral representation:
\be D_{\sigma \sigma}(k^2)= \int ds\, \rho(s)/(k^2 + s -i\epsilon) \ee

One has the sum rule~\cite{akhoury,gunion}
 $\int \rho(s)\, ds = 1$, while otherwise the theory is not renormalizable.
A typical propagator coming from the mixing with a singlet moving
 in $d$ dimensions is of the form~\cite{dilchdok}: 
\be
D_{\sigma \sigma}(q^2)= \left[ q^2 +M^2 - \mu_{lhd}^{8-d}\,\Gamma(-\delta)
(q^2+m^2)^{d-6 \over 2} \right]^{-1} .\ee

In this formula $M$ is the original lagrangian four-dimensional mass of the standard model
field, $m$ is the mass of the higer-dimensional field and $\mu_{lhd}$ is the mass scale that
describes the low-to-high-dimensional mixing interaction. For a detailed derivation from the
lagrangiam we refer to~\cite{bij2006,dilcher}.
This corresponds to a \KL\,\, spectral density in the continuum:

\be  \rho (s) =
 {- (\alpha/ \pi)   \sin( \pi \delta) ( s-m^2)^ { \delta} \over   
[s-M^2 +\alpha \cos(\pi \delta)   ( s-m^2)^ {\delta}]^2
+[\alpha \sin(\pi\delta )   ( s-m^2)^ {\delta}]^2 }.\ee

where we defined 
\be
\alpha = \mu_{lhd}^{8-d}\, \Gamma(-\delta)
\ee
and
\be
\delta = d/2 - 3
\ee
The factor $\Gamma(-\delta)$ is often left out in the literature
and absorbed in the parameters of the propagator. We kept it here
explicitly, so that one can understand formulas for critical dimensions
as a limiting case. 
The construction can be made for all particle types.
For the Higgs scalar there is some tantalizing evidence that this
theory corresponds to reality~\cite{dilcher}.

\section{Unparticles}
In a recent paper~\cite{georgi} a very similar construction was made and called unparticle physics.
Also here there are two sectors of the theory, a singlet sector and the standard model
sector. There are two differences. The singlet sector is supposed to be conformally
invariant and the mixing with standard model particles is through non-renormalizable
instead of renormalizable interactions. The experimental signature is rather similar.
 One finds the exchange of fields with a non-trivial \KL\,\, spectral density.
Due to the assumed conformal symmetry the spectral density[3] is of the form:
\be
\rho(s) \approx s^{d_U-2}
\ee
The following condition was imposed
\be
1<d_U<2.
\ee
Within the very general class of models where an exotic singlet sector is 
coupled to the standard model, imposing such a conformal symmetry is an
extra assumption, that is not strictly necessary. Since ultimately
the interaction with the standard model particles is needed, it is not clear
whether such an assumption can be valid for the full theory. 
The conformal symmetry presumably breaks down at the scale where the
non-renormalizable interactions start playing a role. In the next section we will show
that the unparticle model is a special limit of the HEIDI models. Within
the HEIDI models the conformal symmetry can be studied in detail.
At the same time one can explain the origin of the relation (7).

\section{Comparison}
The starting point is the formula (2) for the propagator or the spectral density
in formula (3). In order to make contact with the unparticle model we
impose conformal invariance in the singlet sector by putting $m=0$.
Now we take the infrared limit $s \rightarrow 0$.  If $d<8$ the spectral density
and the propagator behave in this limit as
\be
\rho(s) \approx s^{3-d/2}
\ee
We therefore make the identification
\be
d_U=5-d/2
\ee
In a somewhat more sophisticated form we can find the same limiting
behaviour by taking $\mu_{lhd} \rightarrow \infty$.
This limit corresponds to a theory with a strong mixing
between the higher-dimensional singlet field and the standard model fields.
It is this particular limit
\be
 m=0,\,\,\,\,\, \mu_{lhd}\rightarrow \infty
\ee
that describes the unparticle. This limit precisely
reproduces the formulas in~\cite{georgi}.
 From the formula (3) for the spectral density it
is clear that for large enough $s$ the exact conformal invariance breaks down,
due to the four-dimensional part containing $(s-M^2)$. For very large $s$ the
density behaves as:
\be
\rho(s) \approx s^{-5+d/2}, \,\,\,\,\,d<8
\ee

\section{Discussion}

We are now in a position to discuss in detail consistency conditions
for the different possibilities. Let us first assume that we are in the
exact conformal model, that is without the $(s-M^2)$ in the density.
For the case $d<8$ the spectral density in the infrared is 
integrable, however not in the ultraviolet. For $d>8$ it is the other way around.
Therefore the completely conformal model is never quite satisfactory.

The situation improves, when we keep the $(s-M^2)$ terms, breaking the 
conformal symmetry due to the presence of standard model fields. However we still
preserve the conformal symmetry in the singlet sector keeping $m=0$. In this 
case the spectral density is integrable in the infrared and the ultraviolet
for all values of $d$. However, if we look at the discrete part of the spectrum
there is always a tachyon pole when $4<d<6$ or $d>8$, indicating an
instability of the model. Only in the range
$6<d<8$ can one avoid a tachyon when $\mu_{lhd}/M$ is small enough.
This explains the origin of the relation (7). So for the range $6<d<8$
one has a propagator that is behaving regularly enough to be seen as
an effective interaction at the tree level.  However the fall-off of
the spectral density at infinity is not fast enough to have a renormalizable theory.
For this one must have $d<6$, but then as we have seen above the higher dimensional
singlet part cannot be conformally invariant and we are back to the original HEIDI models.
At the quantum-mechanical level these appear to be the only consistent theories
of this type.

Some questions~\cite{georgi} raised about the models can be answered.
Is there the possibility for the new sector to have standard model
gauge interactions? In a renormalizable model the answer is clearly no.
Regarding the cosmological consequences it was noted in \cite{bij2006} that
having a higher-dimensional singlet helps in explaining some features of inflation.

\section*{Acknowledgments}
This work was supported by the BMBF Schwerpunktsprogramm
"Struktur und Wechselwirkung fundamentaler Teilchen" and by the
EU network HEPTOOLS. I thank Dr. A. Ferroglia for a careful reading of the manuscript.

\end{document}